\documentclass[12pt,a4paper,final]{iopart}

\usepackage{iopams}  
\usepackage{graphicx}
\usepackage[breaklinks=true,colorlinks=true,linkcolor=blue,urlcolor=blue,citecolor=blue]{hyperref}
\usepackage{color}
\usepackage{url}

\begin{document}

\title[Reduced models for ETG transport in the pedestal]{Reduced models for ETG transport in the pedestal}

\author{D. R. Hatch$^{1*}$,~C. Michoski$^{2}$,~D. Kuang$^{2}$,~B. Chapman-Oplopoiou$^{3}$,M. Curie$^{1}$,~M. Halfmoon$^{1}$,~E. Hassan$^{4}$,~M. Kotschenreuther$^{1}$,~S. M. Mahajan$^{1}$,~G. Merlo$^{3}$,~M. J. Pueschel$^{5,6}$,~J. Walker$^{1}$,~C. D. Stephens$^1$}
\address{$^1$Institute for Fusion Studies, University of Texas at Austin, Austin, Texas, 78712}
\address{$^2$Oden Institute for Computational Engineering and Sciences, University of Texas at Austin, Austin, Texas, 78712}
\address{$^3$CCFE, Culham Science Center, Abingdon OX14 3DB, United Kingdom}
\address{$^4$Oak Ridge National Laboratory, Oak Ridge, Tennessee}
\address{$^5$Dutch Institute for Fundamental Energy Research, 5612 AJ Eindhoven, The Netherlands}
\address{$^6$Eindhoven University of Technology, 5600 MB Eindhoven, The Netherlands}
\ead{$^*$drhatch@austin.utexas.edu}



\begin{abstract}

This paper reports on the development of reduced models for electron temperature gradient (ETG) driven transport in the pedestal.  Model development is enabled by a set of 61 nonlinear gyrokinetic simulations with input parameters taken from the pedestals in a broad range of experimental scenarios.  The simulation data has been consolidated in a new database for gyrokinetic simulation data, the Multiscale Gyrokinetic Database (MGKDB), facilitating the analysis.  The modeling approach may be considered a generalization of the standard quasilinear mixing length procedure.  The parameter $\eta$, the ratio of the density to temperature gradient scale length, emerges as the key parameter for formulating an effective saturation rule.  With a single order-unity fitting coefficient, the model achieves an RMS error of 15\%.  A similar model for ETG particle flux is also described.  We also present simple algebraic expressions for the transport informed by an algorithm for symbolic regression.  
\end{abstract}

\vspace{2pc}
\section{Introduction} \label{introduction}

This paper reports on the development of reduced models for electron temperature gradient (ETG) driven transport in the pedestal.  Reduced models for pedestal transport may facilitate a more comprehensive predictive capability of pedestal structure.  
MHD-based models like EPED have predicted pedestal pressure in many parameter regimes~\cite{snyder_09,snyder_09b}.  However, they typically 
require pedestal-top density and separatrix quantities as inputs and cannot predict pedestal structure in regimes not limited by peeling-ballooning modes~\cite{saarelma_19,frassinetti_20}. 
  Reduced models for pedestal transport may mitigate these weaknesses and expand the scenarios and operating regimes that can be modeled and predicted.  

Reduced models will also faciliate rapid analysis of pedestal transport, thus expanding the number of discharges and scenarios for which pedestal transport can be analyzed and paving the way for real-time analysis.  Moreover, reduced models for ETG may also become a useful complement to the new generation of edge gyrokinetic codes~\cite{ku_09,dorf_18,mandell_20,michels_21}, which are developing comprehensive capabilities for modeling edge turbulence but will likely find the task of brute-force multiscale simulations---spanning the whole range from ion scales to electron scales---beyond even exascale ambitions.  

Recent work has elucidated the instabilities that are most likely responsible for transport in the pedestal in the various transport channels~\cite{kotschenreuther_19,TPT}.  Notably, the disparity between heat diffusivity and particle diffusivity (the latter being much smaller than the former) identified by edge modeling, points toward a vigorous electron heat transport mechanism that needs to be accounted for.  Two instabilities are likely at play: ETG and microtearing modes (MTM)~\cite{hatch_16,hatch_21,hassan_NF_21}.  Since ETG fluctuations exist at scales that are typically inaccessible to diagnostics, we must rely on theory and simulation to infer their activity.  Fortunately, while the smallness
of scale makes them indiscernible to diagnostics, it also makes them more amenable to simulation; the scale separation between ETG scales and background quantities---even in the 
exceedingly narrow pedestal---is sufficient to justify a local flux-tube approach.  There is growing evidence from combined numerical-experimental studies that ETG plays an important role in pedestal 
transport in many H-mode discharges~\cite{told_08,jenko_09,hatch_15,hatch_16,hatch_17,hatch_19,kotschenreuther_19,liu_arxiv_20,chapman_21,guttenfelder_NF_21,TPT}.

Pedestal ETG turbulence is distinct from ETG turbulence in the core.  The extreme density and temperature gradients in the pedestal far surpass those of the 
background magnetic field, thus circumventing the typical magnetic drift resonances and favoring slab resonances (for an exception to this claim, 
see Refs.~\cite{told_08,parisi_NF_20}, which identified toroidal ETG modes destabilized at large radial wavenumbers).  This results in 
turbulence (1) that is isotropic in comparison with the streamer-dominated core ETG~\cite{jenko_00b,dorland_prl_00,nevins_06}; (2) exhibits high-$k_z$ structure, which demands extreme resolution 
in the parallel direction; and (3) has contributions from a high number (10-20) of unstable eigenmodes at each wavenumber.  Properties (2) and (3) challenge some of the 
standard approaches to reduced quasilinear modeling, making the present work challenging and timely.

In this paper, we exploit a newly-created database for gyrokinetic simulation data---the Multiscale Gyrokinetic Database (MGKDB)---in order to 
formulate reduced models for pedestal transport from ETG turbulence.  MGKDB is a community resource
for storing and analyzing gyrokinetic and reduced model simulations that have been run over the
last decades.   It utilizes a non-relational MongoDB \cite{banker2016mongodb} (NoSQL) based structure to maximize flexibility so that any output format from any
particular code infrastructure can be easily supported.  In addition, MGKDB seeks compatibility with an international IMAS data standard \cite{romanelli2020code} to containerize its quantities of interest and interfaces these data types with a comprehensive python library called the Ordered Multidimensional Array Structures (OMAS) library \cite{Meneghini_2020}, that allows for easy conversion to other data formats including, for example, SQL-based formats.  MGKDB may be accessed remotely through either python scripts, command shell options (i.e. MongoDB or Python), a custom graphical user interface, or existing MongoDB GUIs.

The simulations of which this database is comprised were performed with the GENE gyrokinetic code~\cite{jenko_00b} applied to
multiple radial locations in the pedestal for discharges spanning multiple devices (DIII-D, JET, C-Mod, AUG) and operating scenarios.  Most of the simulations have been previously described 
in at least one of Refs.~\cite{hatch_15,hatch_16,hatch_17,hatch_19,kotschenreuther_19, hassan_pop_21,chapman_21,liu_arxiv_20,TPT}. 

We pursue two general approaches to reduced modeling.  First, we investigate variations on the standard quasilinear mixing length approach, wherein a turbulent diffusivity is approximated with step size determined by the perpendicular wavelength of the eigenmode and step time by the linear growth rate.  Our modest variation entails allowing for additional parameter dependences in the saturation rule, which are guided by the dataset of nonlinear simulations.  Second, we formulate simple algebraic expressions for the transport using a symbolic regression algorithm.  In both approaches, the parameter $\eta = L_n/L_{Te}$, the ratio of the density to temperature gradient scale length, emerges as the key parameter.  This is consistent with recent theoretical~\cite{hatch_17,hatch_19} and experimental~\cite{frassinetti_21} studies identifying the importance of this parameter for the JET pedestal.    

The paper is outlined as follows.  The MGKDB database and the dataset are described in Sec.~\ref{sec:dataset}.  The reduced model based on a quasilinear
mixing length estimate is described in Sec.~\ref{sec:quasilinear}.  Various analytic models are described in Sec.~\ref{sec:analytic}.  Summary and conclusions are found in 
Sec.~\ref{sec:summary}.  

\section{MGKDB and The Data Set}
\label{sec:dataset}

The dataset consists of 61 nonlinear single-scale ETG simulations.  The main parameters for these simulations are shown
in the table in~\ref{appendixA}.  The data is shown visually in Fig.~\ref{QNL_vs_parameters}, which plots the gyroBohm-normalized nonlinear heat flux.  Here, the gyroBohm heat flux is defined as $Q_{GB}=n_0 T_0 c_s \frac{\rho_s^2}{a^2}$, where $\rho_s=c_s/\Omega_i$ is the sound gyroradius, $c_s=\sqrt{T_{0e}/m_i}$ is the sound speed, 
$\Omega_i= e B_0/m_i$ is the ion gyrofrequency, $n_{0e},T_{0e}$ are the background electron density and temperature, and $a$ is the minor radius.  The parameter $\langle k_y \rangle$
is the spectrally-weighted value of $k_y$, i.e., a value of $k_y$ that is representative of the nonlinearly-saturated turbulence and not an input parameter.  Most of 
these simulations were carried out while studying actual discharges on JET, DIII-D, AUG, and C-Mod.  Many of the simulations are scans across radial positions in the 
pedestal or variations of background gradients within error bars.  Most of the simulations are described in the following recent publications: 
Refs.~\cite{hatch_15,hatch_16,hatch_17,hatch_19,kotschenreuther_19, hassan_pop_21,chapman_21,liu_arxiv_20,TPT}.
  All simulations were uploaded to the MGK database (MGKDB), which was exploited for the analysis in this work.   

Many (but not all) of the simulations were subjected to extensive convergence tests.  Generally, for a given study, convergence tests were carried out for a base case 
and subsequent scans retained the nominal parameter setup.  
All simulations were examined to ensure the following: (1) saturated heat fluxes, and (2) well-behaved (heat flux peaking much higher than the minimum wavenumber and substantial fall off at high k) heat flux spectra in the perpendicular wavenumbers.       

Forty-eight of the simulations employ an adiabatic ion approximation and thus dynamically evolve only the electron species.  The inclusion of kinetic ions generally does not produce qualitative differences in the heat flux (ion dynamics are strongly suppressed by FLR effects at these electron scales)~\cite{ hatch_15,hatch_16,hatch_17}.  Six simulations include both ions and electrons and seven more simulations include three species (including a dynamic impurity).  The main transport channel for ETG modes is electron heat.  Particle transport is enforced to be zero for simulations with adiabatic ions.  Even with kinetic ions, particle transport is generally weak due to FLR suppression of kinetic ion dynamics, which in turn suppresses particle flux via ampibolarity.  Nonetheless, particle sources are also generally weak in the pedestal, so even low levels of ETG particle flux may be relevant~\cite{TPT}.  Consequently, we use the thirteen kinetic-ion simulations to
generalize the model to include particle flux.  

As is common for simulation studies, some potentially-important effects have been neglected.  Most notably, we have not accounted for potential 
multiscale effects.  Ref.~\cite{pueschel_20} identified, for an idealized setup targeting pedestal-relevant parameters, a reduction of ETG transport due to interaction with 
ion-scale microtearing turbulence (or, rather, zonal flows stimulated by it).  There is also a possibility of multiscale interaction between different branches of ETG: slab ($k_y \rho_s \sim 100$) and toroidal~\cite{parisi_NF_20} ($k_y \rho_s \sim 10$).  Although a rigorous 
survey of multiscale effects lies beyond the scope of this work, some simulations were spot-checked with the goal of probing the effects of toroidal ETG modes.  
For these cases, we did not identify large heat fluxes from toroidal ETG modes nor did interaction with toroidal ETG modes significantly alter the transport levels from slab ETG modes.  However, we acknowledge the possibility that such dynamics may play a role for yet-better-resolved and/or longer-simulated runs and/or parameter points that we did not investigate. 

We also note that many of the simulations in our dataset produce transport levels in close proximity to the experimental expectations (as noted in previous publications~\cite{hatch_15,hatch_16,hatch_17,hatch_19,kotschenreuther_19, hassan_pop_21,chapman_21,liu_arxiv_20,TPT}), and none of the simulations significantly exceed experimental transport levels.  In short, although uncertainties remain, we consider it very likely that the simulations in this database represent realistic predictions of pedestal ETG transport and that ETG transport often plays a significant role in the pedestal power balance. 

\begin{figure}[htb!]
    \centering
    \includegraphics[scale=0.8]{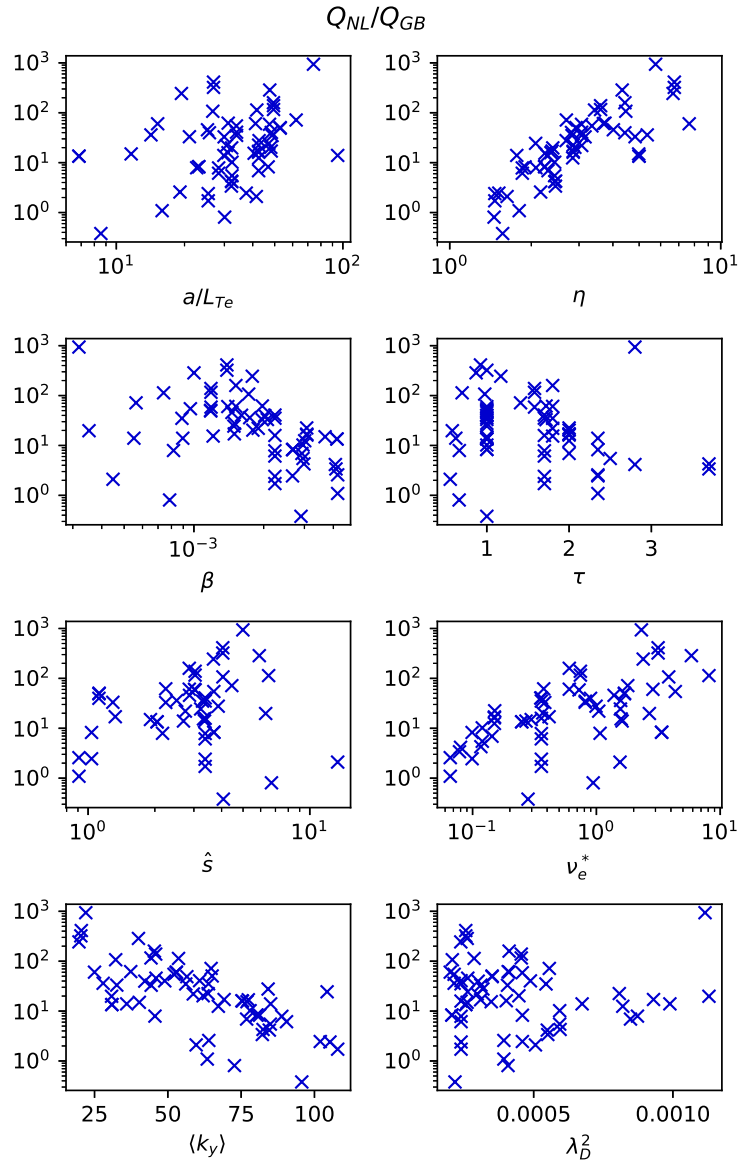}
    \caption{\label{QNL_vs_parameters} GyroBohm-normalized heat flux from nonlinear simulations in the data set plotted against several parameters.  This data set is used to formulate and test the reduced models in this paper.  All parameters are simulation inputs except $\langle k_y \rangle$, which represents the peak wavenumber of the nonlinear heat flux spectrum as defined in the text.  }
\end{figure}

\section{Generalized Quasilinear Models}
\label{sec:quasilinear}

We first investigate standard quasilinear~\cite{staebler_05,bourdelle_07,citrin_17,stephens_21} mixing-length approaches, and generalizations thereof, as illustrated in Eq.~\ref{eq:generalized_mixing}.
\begin{equation}
    \label{eq:generalized_mixing}
    Q_{QL} = a_0({\bf \Theta}) \omega_{Te} \textrm{MAX}_{ky} \left \{ \frac{\gamma}{\langle k_{\perp}^2 \rangle} \right \} .
\end{equation}
Here, $Q$ is the electron electrostatic heat flux (gyroBohm normalized as defined above), $\omega_{Te}=\frac{a}{L_{Te}} = \frac{1}{T_e}\frac{dT_e}{d\rho_{tor}}$ is the normalized inverse electron temperature gradient scale length ($\rho_{tor}$ is the square root of the normalized toroidal magnetic flux).  The variable $a_0$ is a fitting parameter and ${\bf \Theta} = \left ( \omega_{Te},\omega_{ne},\beta,\hat{s},\tau,\lambda_D,\nu^*_E \right )$ simply denotes the possibility of incorporating additional parameter dependences into the saturation rule.  Here, $\beta$ is the ratio of thermal to magnetic energy density, $\tau = \frac{T_{e0}}{T_{i0}} Z_{eff}$, $Z_{eff}=\frac{1}{n_e}\sum_j Z_j^2 n_j$, $\lambda_D$ is the Debye length, $\nu^*_e$ is the normalized electron collision frequency, $\omega_{ne}=\frac{a}{L_{Te}} = \frac{1}{n_e}\frac{dn_e}{d\rho_{tor}}$ is the normalized inverse electron density gradient scale length, and $\hat{s} = \frac{\rho_{tor}}{q}\frac{dq}{d \rho_{tor}} $ is the magnetic shear.  These quantities are listed and defined in~\ref{appendixA}.     

A scan over $k_y$ of linear gyrokinetic simulations is used to formulate the mixing length estimate, 
$\gamma/\langle k_\perp \rangle^2$, where $\gamma$ is the linear growth rate (normalized to the ratio of the sound speed to minor radius $c_s/a$) and the eigenmode-averaged perpendicular 
wavenumber is
\begin{equation} 
    \label{eq:kperp2_eigenmode_average}
    \langle k_\perp^2  \rangle = \frac{\int k_\perp^2 |\phi|^2 \kappa(k_\perp) dz}{\int |\phi|^2 \kappa(k_\perp) dz}.  
\end{equation}
In this equation, $z$ is the distance along the field line, parameterized by the poloidal angle (our standard domain for this problem is $-7 \pi<z<7 \pi$; smaller integration ranges were also tested with negligible difference), $\phi$ is the electrostatic
potential for the eigenmode, $k_\perp^2 = g_{xx} k_x^2 + 2 g_{xy} k_x k_y + g_{yy} k_y^2$ is the perpendicular wavenumber, 
$k_x(k_y)$ is the radial (binormal) wavenumber, $g_{i j}$ are the relevant components of the metric tensor, and $\kappa(k_\perp) = \left (1 + \frac{2 ( k_\perp^2 + 2/3 \pi k_\perp^4)  }{(1 + 2/3 k_\perp^2)  } \right )^{-1/2} \approx J_0(k_\perp)^2$ approximates the Bessel functions that represent gyroaveraging.  Note that, 
due to magnetic shear, the radial wavenumber is connected to the parallel coordinate, $z$, as follows: $k_x = z \hat{s} k_y$.  All wavenumbers are normalized to the sound gyroradius $\frac{m_i c_s}{e B_0}$.  

The mixing length $\gamma / \langle k_\perp^2 \rangle$ is motivated as a turbulent diffusivity with relevant scale length,
$1/k_\perp$, and time scale, $1/\gamma$, set by the linear instabilities.  We chose the scan range $ 10 \leq k_y \rho_s \leq 240$.  
The maximum value of the mixing length diffusivity over the scan is selected for the model.  Several variations were considered, and, to 
some extent, tested, including: (1) scanning also the ballooning angle, (2) maximizing or summing over multiple eigenmodes using the 
GENE eigenmode solver, (3) summing (as opposed to maximizing) over the $k_y$ scan, and (4) including an additional 
factor of the ratio of the heat flux to the density fluctuation amplitude $Q/|n|^2$.  None of these generalizations 
substantially improved the model and some actually reduced accuracy.  Consequently, we have retained the simplest, and 
most computationally inexpensive, approach: limiting the scan to the most unstable eigenmode at zero ballooning angle and 
maximizing the mixing length over $k_y$.  

 
\subsection{Heat Flux}
\label{sec:heat_flux}
 
\begin{figure}[htb!]
    \centering
    \includegraphics[scale=0.8]{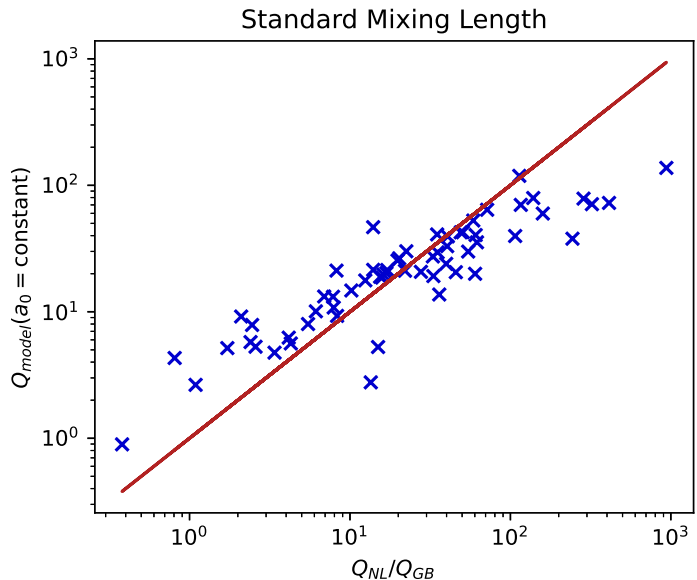}
    \caption{\label{fig:QL_naive} The standard mixing length estimate (assuming $a_0$ is constant) (y axis) plotted against the nonlinear simulations result (x axis) for the same parameter point.  As with many following figures, the accuracy of the model can be gauged by how closely the points cluster around the line.  The error for this model is $\varepsilon = 0.35$.}
\end{figure}

The assumption of constant $a_0$ represents the standard mixing length estimate, which has been effective in modeling transport 
in several scenarios, including Refs.~\cite{bourdelle_07,merz_08,xie_20}.  We test this simplest expression right away 
and find that it results in substantial errors when attempting to model the transport across the dataset, as shown in Fig.~\ref{fig:QL_naive}.

To quantify the accuracy, we define a modified relative error as
\begin{equation}
   \varepsilon =  \sqrt{\frac{1}{N} \sum \frac{\left ( Q_{NL} - Q_{model} \right )^2}{ ( Q_{model} + Q_{NL})^2} }, 
\label{error}
\end{equation}
which equally penalizes errors in the limits $Q_{model} \ll Q_{NL}$ and $Q_{model} \gg Q_{NL}$ ($N$ is the number of simulations in the dataset).  The standard mixing length estimate produces $\varepsilon = 0.35 $.  


\begin{figure}[htb!]
    \centering
    \includegraphics[scale=0.8]{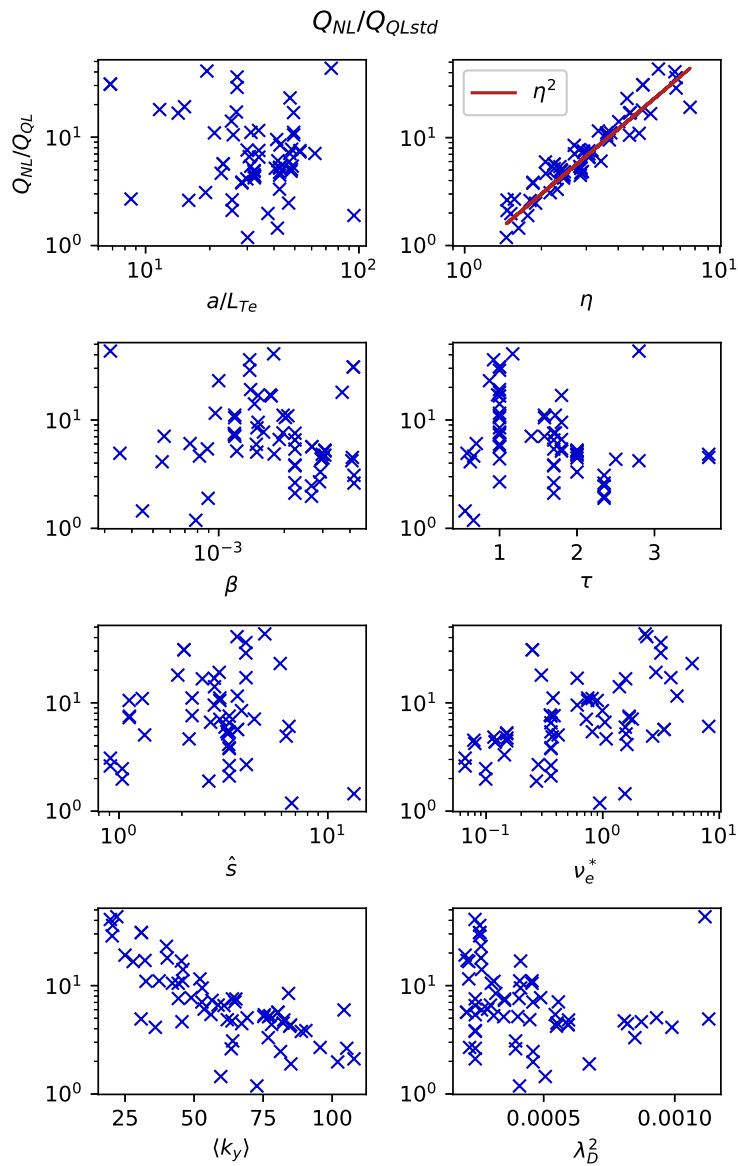}
    \caption{\label{fig:QNL_over_QLL_vs_parameters} The ratio of the heat flux to the standard mixing length estimate is plotted against the parameters in order to identify additional relevant parameter dependences for the saturation rule.  A clear correlation with $\eta^2$ is identified as shown by the line.}
\end{figure}

\begin{figure}[htb!]
    \centering
    \includegraphics[scale=0.8]{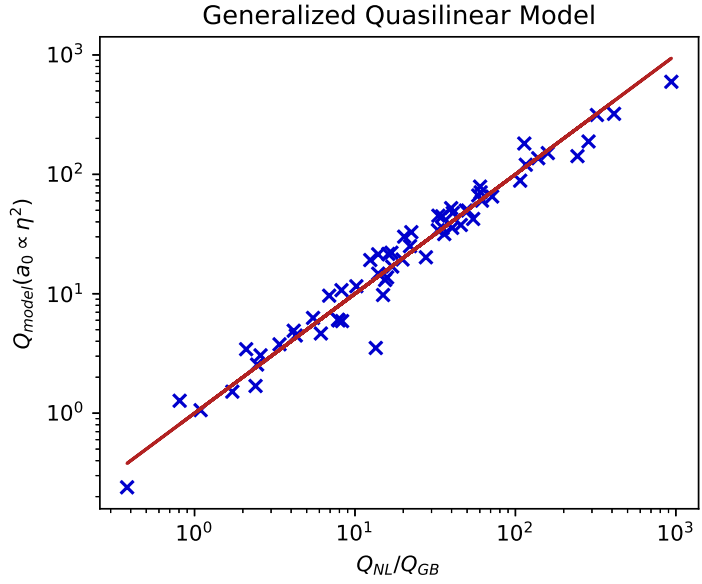}
    \caption{\label{fig:eta2_QQL} The quasilinear model defined in Eq.~\ref{eq:eta2_QQL} is quite accurate ($\varepsilon = 0.15$) and represents the major result of this work.   }
\end{figure}

As a next step, we return to the database to investigate additional parameter dependences in 
$a_0 = \frac{Q_{NL}}{\omega_{Te}\textrm{MAX}_{ky} \left \{ \frac{\gamma}{\langle k_{\perp}^2  ]} \right \}}$, 
which we define here as the ratio between the nonlinear heat flux and the mixing length estimate.  This is shown 
in Fig.~\ref{fig:QNL_over_QLL_vs_parameters}, where one clear correlation immediately appears: 
$a_0 \propto \eta^2$.  We modify this slightly as follows with an eye toward future application to scenarios with extremely 
weak density gradients: $\eta = \frac{\omega_{Te}}{\omega_{ne}} \rightarrow \hat{\eta} = \frac{\omega_{Te}}{1+\omega_{ne}}$.  This modification has very little effect 
within the current dataset, for which $\omega_n$ is typically much larger than unity, but ensures well-behaved solutions in 
the limit $\omega_n \rightarrow 0$.  We thus arrive at the model that constitutes the core result of this paper:

\begin{equation}
\label{eq:eta2_QQL}
Q_{QL} = 0.87 \hat{\eta}^2 \frac{a}{L_{Te}} \textrm{MAX}_{ky} \left \{ \frac{\gamma}{\langle k_{\perp}^2 \rangle} \right \} ,
\end{equation}
which is shown in Fig.~\ref{fig:eta2_QQL}.

As can be seen in Fig.~\ref{fig:eta2_QQL}, with few exceptions, the model accurately recovers the nonlinear heat flux across the 
dataset.  The error for the model defined in Eq.~\ref{eq:eta2_QQL} is $\varepsilon = 0.15$ (for reference, the error for the standard quasilinear model is $\varepsilon = 0.35$).
 We, thus, have arrived at a model based on a physical (gyrokinetic) quasilinear mixing-length estimate, 
one additional parameter dependence ($\eta^2$), and a single order-unity fitting parameter, that effectively reproduces 
heat fluxes from a large dataset of nonlinear gyrokinetic simulations.   

\begin{figure}[htb!]
\centering
\includegraphics[scale=0.8]{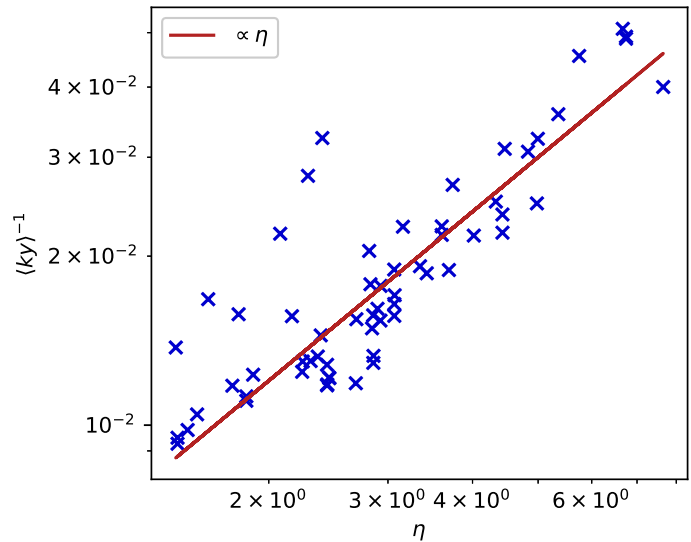}
\caption{\label{fig:kyavg_vs_eta}  The inverse of the spectrally-averaged wavenumber $\langle k_y \rangle$ plotted against $\eta$.  }
\end{figure}

While a thorough investigation of the origin of the additional $\eta$ dependence is beyond the scope of this work, we will make a few simple observations.  Fig.~\ref{fig:kyavg_vs_eta} shows the spectrally-averaged $k_y$ plotted against $\eta$, demonstrating a clear proportionality between the two.  The nonlinear spectrum condenses at lower wavenumbers as $\eta$ increases, thus enhancing the transport.  There is some evidence in the dataset that this downshift exceeds that predicted by the mixing-length estimate, thus requiring the additional factor of $\eta^2$ to compensate.  A deeper understanding of the $\eta^2$ dependence will be pursued in future work.

\subsection{Particle Flux}

In this section, we investigate similar reduced models for particle transport from pedestal ETG.  Particle flux from ETG modes is 
very modest; the ambipolar (net zero radial charge flux) nature of gyrokinetic transport 
means that electron particle transport is constrained to small levels by the low ion particle transport due to FLR effects 
on the ions.  However, the pedestal parameter regime of interest is also characterized by low particle transport; several studies have demonstrated via edge modeling that electron heat diffusivity greatly surpasses electron particle diffusivity 
in the pedestal~\cite{callen_10,TPT}: $D_e / \chi_e \ll 1$.  Gyrokinetic simulations have demonstrated that, while ETG 
particle transport is not likely to account for the bulk of the transport, it is often at levels that are not negligible~\cite{TPT}.  
Moreover, gyrokinetic simulations often exhibit particle pinches~\cite{TPT}, which could be important for fueling the pedestal beyond the 
capacity of neutral penetration alone.  

\begin{figure}[htb!]
    \centering
    \includegraphics[scale=0.8]{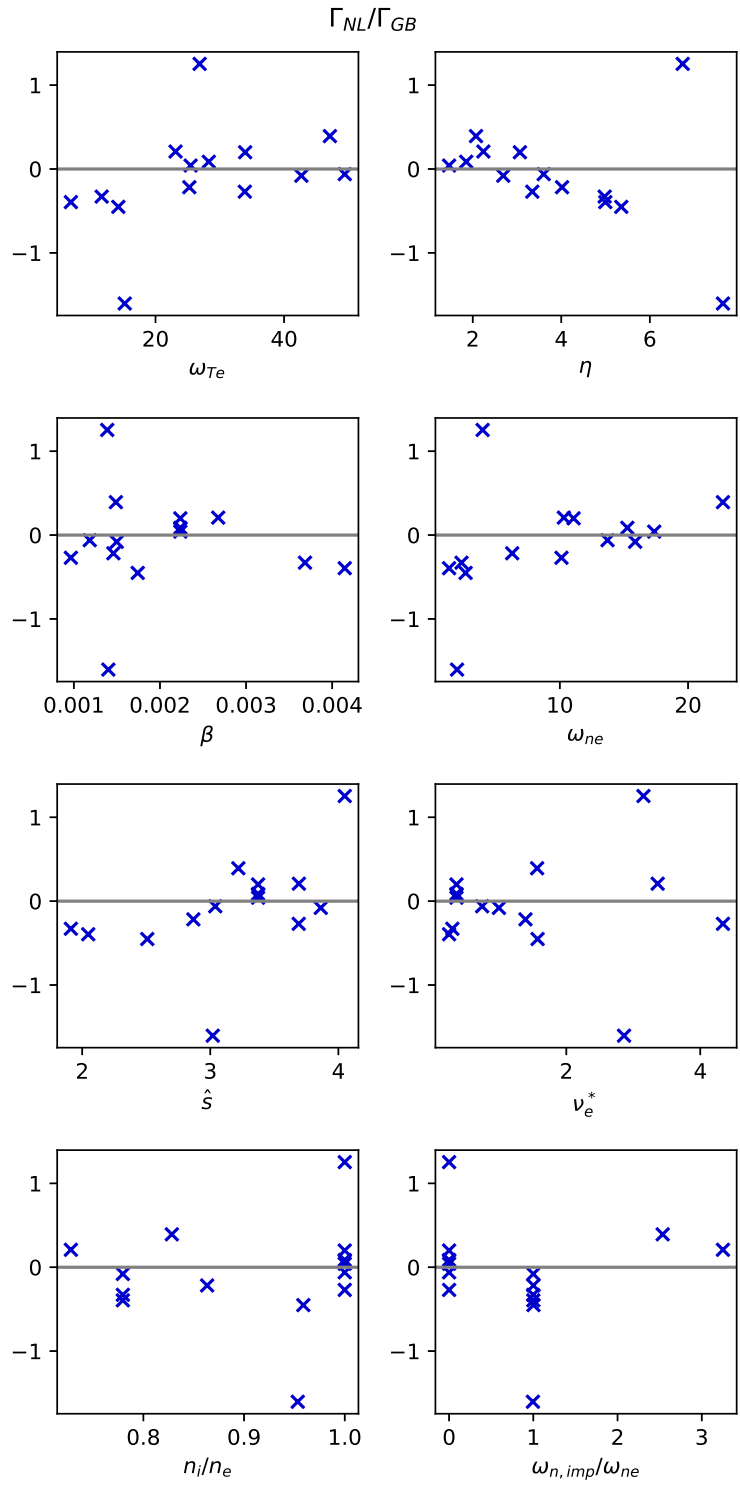}
    \caption{\label{fig:Gamma_vs_parameters} The gyroBohm-normalized particle flux plotted against several parameters.}
\end{figure}

Thirteen simulations in the dataset retain kinetic ions (seven of those including also an impurity species).  Kinetic ions also open the possibility of additional instabilities in the lower wavenumber ranges, but such instabilities have not been observed in these scenarios for the wavenumber ranges of interest.  The electron particle
flux for these simulations is shown in Fig.~\ref{fig:Gamma_vs_parameters} 
(normalized to $\Gamma_{GB} = \frac{n_0 \rho_s^2 c_s}{a^2}$).  Note the existence of both positive and negative 
fluxes in the dataset and that the density gradient $\omega_n$ roughly parameterizes the transition between the two signs.  
Two simulations (at high $\eta$) exhibit particle fluxes of order unity, one positive and one negative.  The simulation with 
a large positive particle flux is particularly anomalous considering its relatively low density gradient and its  
relative close proximity to the large negative simulation for most parameters.  

\begin{figure}[htb!]
    \centering
    \includegraphics[scale=0.8]{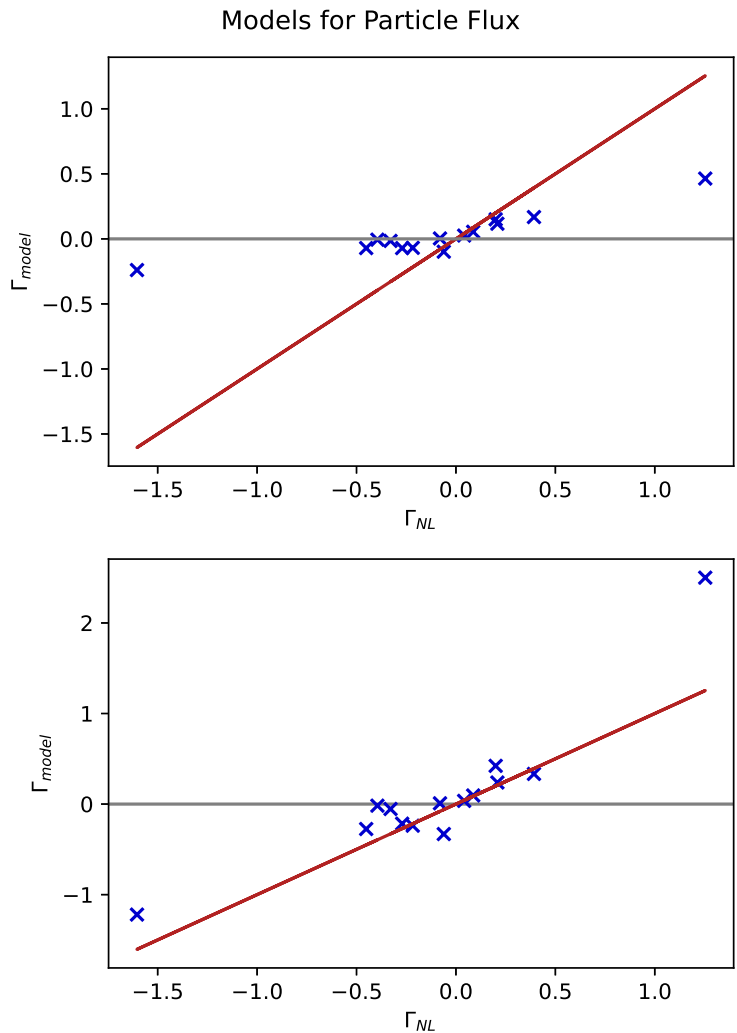}
    \caption{\label{fig:Gamma_QL}  The models  $\Gamma_{QL,naive} = Q_{QL} \textrm{MAX}_{ky}\left \{ \frac{\Gamma_e}{Q_e} \right \}$ (Eq.~\ref{eq:gammaQLnaive}) (top) and $\Gamma_{QL} = \hat{\eta} Q_{QL} \textrm{MAX}_{ky}\left \{ \frac{\Gamma_e}{Q_e} \right \}$ (Eq.~\ref{eq:gammaQL}) (bottom).  Note that Eq.~\ref{eq:gammaQL} successfully reproduces the sign of the particle flux for nearly all (one exception) simulations. }
\end{figure}

Perhaps the most obvious generalization of the model to include particle flux would weight the heat flux prediction (Eq.~\ref{eq:eta2_QQL}) by the quasilinear ratio of the fluxes as follows:
\begin{equation}
    \label{eq:gammaQLnaive}
    \Gamma_{QL,naive} = Q_{QL} \textrm{MAX}_{ky}\left \{ \frac{\Gamma_e}{Q_e} \right \} ,
\end{equation}
where $Q_{QL}$ is the model defined above for the heat flux (defined in Eq.~\ref{eq:eta2_QQL}), and the final term denotes the maximum of the ratio of the particle to heat flux defined by the linear modes maximized over the $k_y$ scan.  As shown in Fig.~\ref{fig:Gamma_QL}, this model is not very accurate, under predicting in particular the extreme flux cases.  We find that one additional factor of $\eta$ improves the model substantially: 
\begin{equation}
    \label{eq:gammaQL}
    \Gamma_{QL} = \hat{\eta} Q_{QL} \textrm{MAX}_{ky}\left \{ \frac{\Gamma_e}{Q_e} \right \} ,
\end{equation}
  Note that this model includes one \textit{additional} factor of $\eta$ beyond those already included in $Q_{QL}$.  As seen in Fig.~\ref{fig:Gamma_QL}, this improves the agreement significantly, particularly for the cases with large fluxes.  Notably, this model reproduces the transition between positive 
and negative fluxes and distinguishes between the two simulations with large fluxes (one positive and one negative) despite 
their apparent proximity in parameter space.

\section{Algebraic Expressions}
\label{sec:analytic}

Although we have formulated a rather accurate quasilinear model for pedestal transport, this model still requires several linear gyrokinetic simulations.  This represents enormous savings in comparison with full nonlinear simulations, but a simple algebraic expression would still be desirable for the purposes of rapid evaluation and physical intuition.  Consequently, as a final investigation, we abandon the quasilinear mixing-length framework entirely and investigate simple algebraic expressions for the fluxes.  

To this end, we apply a novel symbolic regression algorithm, which
minimizes the error by systematically surveying combinations of
pre-selected algebraic forms.  More specifically, the algorithm,
called System Identification and Regression (SIR), minimizes Eq.~\ref{optm:rational} relative to a collection of rational functions of fixed
top admissible monomial degree $d$ (in its elements), fixed top
admissible nonlinearity order $n_{\ell}$ per input (in its elements),
and maximum number of linearly combined terms $n_{u}$ allowed in the
resulting expression (in either the numerator or the denominator).
For example,
\begin{equation}\label{optm:rational}
   \mathrm{MIN}_{a,b}\quad \varepsilon \left(\boldsymbol{y}, \quad
\frac{a_1\mathbf{P}_1(\boldsymbol{u})+\ldots+a_m\mathbf{P}_m(\boldsymbol{u})}{b_1\mathbf{Q}_1(\boldsymbol{u})+\ldots+b_n\mathbf{Q}_n(\boldsymbol{u})}\right),
\end{equation}
where $\boldsymbol{y}$ is the target, $\boldsymbol{u}$ are the inputs,
and $a_i$ and $b_i$ are the coefficients on the monomial numerator
$\mathbf{P}_i$
and denominator $\mathbf{Q}_i$ terms.  The resulting $\varepsilon$ is
then computed and stored
at each degree $d_{0}\in [0,d]$ consecutively for all admissible
nonlinear combinations starting at $n_u=1$, and results in an ordered
set of candidate rational function expressions effectively minimizing
$\varepsilon$.  These candidate expressions are then surveyed, and the
most likely resultant expression is chosen based on physics insight.

In carrying out this exercise, we are wary of over-fitting, particularly for the small set of particle flux data.  
Consequently, we favor simple expressions with a willingness to sacrifice to some extent accuracy.  

\begin{figure}[htb!]
    \centering
    \includegraphics[scale=0.8]{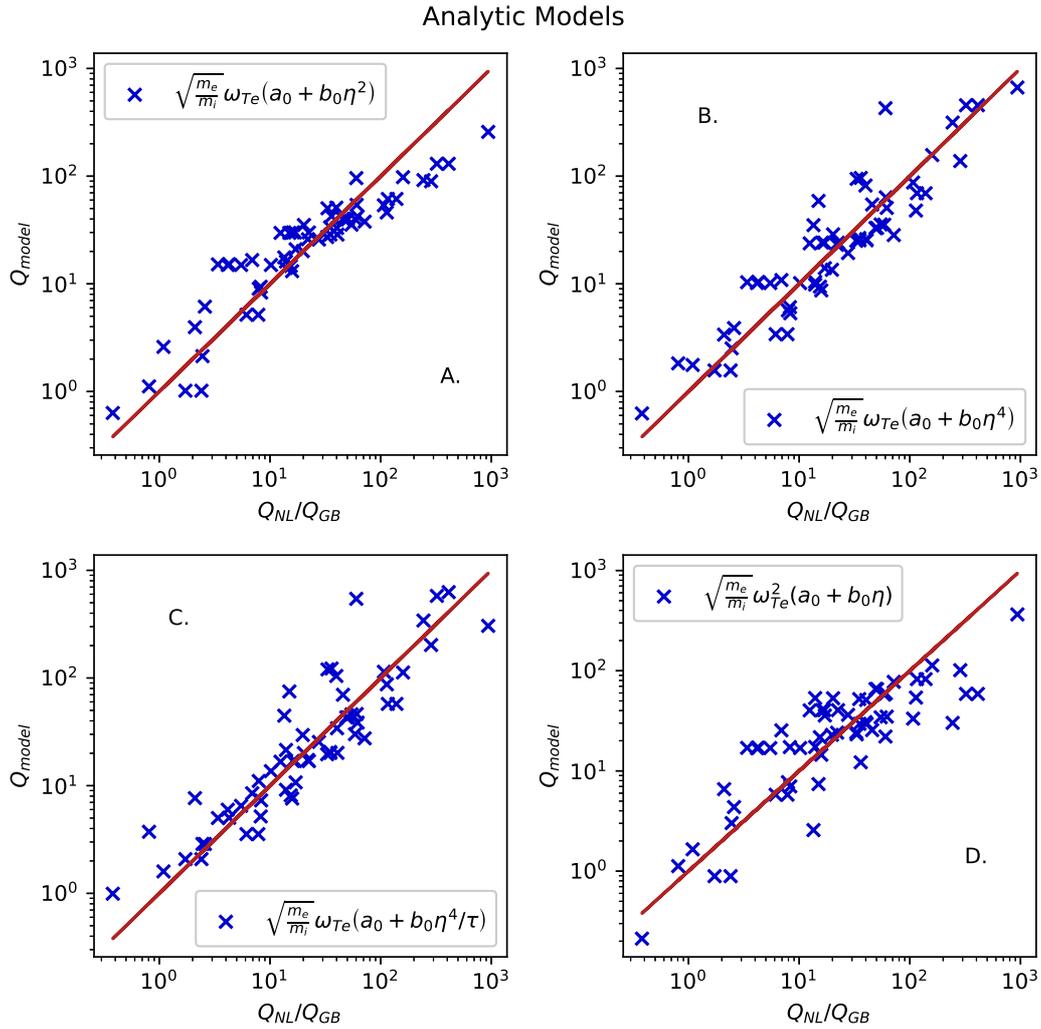}
    \caption{\label{fig:Q_analytic} The models defined in Eqs.~\ref{eq:SIR1},~\ref{eq:SIR2},~\ref{eq:SIR3},~\ref{eq:guttenfelder} plotted in A-D respectively.}
\end{figure}

Fig.~\ref{fig:Q_analytic} shows four models defined in the following equations (corresponding to A-D respectively):
\begin{equation}
    \label{eq:SIR1}
    Q_1 = \sqrt{\frac{m_e}{m_i}}\omega_{Te} \left (a_0 + b_0 \eta^2 \right )
\end{equation}
with $a_0=-12.1$ and $b_0 = 6.73$, with error $\varepsilon=0.290$, 
    
\begin{equation}
\label{eq:SIR2}
Q_2 = \sqrt{\frac{m_e}{m_i}}\omega_{Te} \left (a_0 + b_0 \eta^4 \right )
\end{equation}
with $a_0=1.44$ and $b_0 = 0.50$ with error $\varepsilon=0.279$,

\begin{equation}
\label{eq:SIR3}
Q_3 = \sqrt{\frac{m_e}{m_i}}\omega_{Te} \left (a_0 + b_0 \eta^4/\tau \right )
\end{equation}
with $a_0=3.23$ and $b_0 = 0.63$, with error $\varepsilon=0.303$, 

\begin{equation}
    \label{eq:guttenfelder}
    Q_4 = \sqrt{\frac{m_e}{m_i}}\omega_{Te}^2 \left (a_0 + b_0 \eta \right )
\end{equation}
with $a_0 = -1.26$, and $b_0 = 0.919$, with error $\varepsilon = 0.368$.


The final expression follows the form proposed in Ref.~\cite{guttenfelder_NF_21}.  This does indeed qualitatively capture the major trend of the data but is not as accurate for this data set as Eqs.~\ref{eq:SIR1},\ref{eq:SIR2},\ref{eq:SIR3}.  We view Eqs.~\ref{eq:SIR2},~\ref{eq:SIR3} as likely the most reliable.  Eq.~\ref{eq:SIR2} captures the major trends using only simple combinations of the gradients.  Eq.~\ref{eq:SIR3} additionally incorporates a factor of $\tau =  \frac{T_{e0}}{T_{i0}}Z_{eff}$.  Note that $\tau$ is only relevant for simulations with adiabatic ions; it captures the effects of the ions in the field equation and is well known to be stabilizing.  

Many similar expressions produce similar accuracy.  For example, in the following expression, the model applies the exponent outside the parentheses in a form that would reflect threshold behavior more transparently:
\begin{equation}
    \label{eq:Q5}
    Q_5 = \sqrt{\frac{m_e}{m_i}} a_0\omega_{Te} \left (b_0 + \eta \right )^4
\end{equation}
with $a_0 = 0.309$, $b_0 = 0.413$, and error $\varepsilon = 0.271$.  The result is very similar to that in Eq.~\ref{eq:SIR3}. Slight differences like these, however, may become important when attempting to capture the transport near threshold.
 
Although we have used a very simple gyroBohm normalization (defined above), the more-natural variation for ETG transport would be~\cite{hatch_15,guttenfelder_NF_21} $Q_{eGB} = n_{e0} T_{e0} v_{Te} \frac{\rho_e^2}{ L_{Te}^2 } $, where $L_{Te}$ is the electron temperature gradient scale length.  If this is interpreted in terms of a Fick's law $Q = n \nabla T \chi$, then one factor of $1/L_{Te}$ comes from the gradient and the other comes from the assumption that the ETG growth rate scales like $v_{Te}/L_{Te}$.  However, for slab ETG, the growth rate is dependent on $\eta=L_n/L_T$ as opposed to $L_{Te}$ alone.  This may explain the superior fit in Eqs.~\ref{eq:SIR1},~\ref{eq:SIR2},~\ref{eq:SIR3} (which entail a single factor of $\omega_{Te}$)  in comparison with Eq.~\ref{eq:SIR2} (with $\omega_{Te}^2$).

We note also note that Ref.~\cite{jenko_04} proposes a diffusivity for ETG transport with very strong dependence on gradient scale lengths: $Q_e \propto \omega_{Te}^5$ similar to Eqs.~\ref{eq:SIR2},\ref{eq:SIR3}.  However, in contrast to our expressions, the model includes only temperature gradients but not density gradients (i.e., it is not parameterized in terms of $\eta$ and does not capture the stabilizing effects of density gradients).  This may be attributable to its focus on a core-like parameter regime, where curvature-driven (as opposed to slab) ETG is salient.

\begin{figure}[htb!]
    \centering
    \includegraphics[scale=0.8]{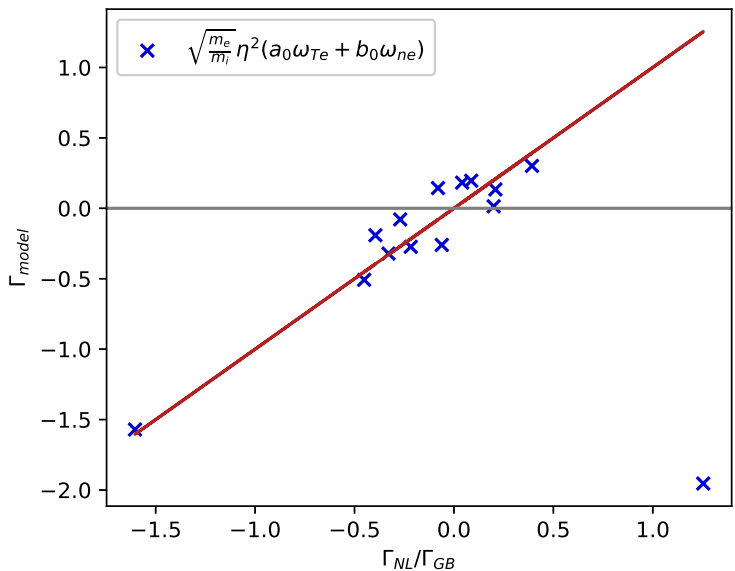}
    \caption{\label{fig:analytic_particle} The model defined in Eq.~\ref{eq:SIR4particle}.}
\end{figure}

For particle transport, we propose an expression that includes a diffusive and pinch component, amplified by $\eta^2$ as shown in Fig.~\ref{fig:analytic_particle} and defined in Eq.~\ref{eq:SIR4particle}. 
\begin{equation}
    \label{eq:SIR4particle}
    \Gamma_1 = \sqrt{\frac{m_e}{m_i}}  \eta^2 \left (a_0 \omega_{Te}  + b_0 \omega_{ne} \right )
\end{equation}
with $a_0 = -0.18$ and $b_0 = 0.56$.
Note that these expressions (as well as all others we investigated) is not capable of capturing the simulation with large positive particle flux, in contrast with the quasilinear 
model defined above in Eq.~\ref{eq:gammaQL}.  It is likely that a more sophisticated treatment of geometry and/or impurities would be necessary to achieve a more comprehensive expression for the particle flux.

\section{Summary and Conclusions}
\label{sec:summary}
This paper has presented reduced models for ETG transport in the pedestal.  The development of the models exploited a dataset of 61 nonlinear simulations from the MGKDB database.  As may be expected for slab ETG modes, the parameter $\eta$ emerges as the key parameter for both the quasilinear mixing-length approach as well as simple algebraic expressions for the transport.  The most important models are reproduced here for easy reference.    

The best quasilinear mixing length model for the heat flux identified in this work ($\hat{\eta} = \omega_{Te}/(1+\omega_{ne})$):
\begin{equation}
Q_{QL} = 0.867 \hat{\eta}^2 \omega_{Te} \textrm{MAX}_{ky} \left \{ \frac{\gamma}{\langle k_{\perp}^2 \rangle} \right \} 
\end{equation}

The best quasilinear mixing length model for the particle flux identified in this work:
\begin{equation}
    \Gamma_{QL} = \hat{\eta} Q_{QL} \textrm{MAX}_{ky}\left \{ \frac{\Gamma_e}{Q_e} \right \} ,
\end{equation}
where $Q_{QL}$ is defined immediately above.

The best algebraic expression for the heat flux identified in this work (un-normalized with quantities defined in Sec.~\ref{sec:dataset}):
\begin{equation}
Q_3 = \left [ \sqrt{\frac{m_e}{m_i}} n_{0e} T_{0e} c_s \frac{ \rho_s ^2}{ a^2} \right ] \omega_{Te} \left (3.23 + 0.63 \eta^4/\tau \right )
\end{equation}
or
\begin{equation}
Q_2 =  \left [ \sqrt{\frac{m_e}{m_i}} n_{0e} T_{0e} c_s \frac{ \rho_s ^2}{ a^2} \right ] \omega_{Te} \left (1.44 + 0.5 \eta^4 \right ).
\end{equation}
Slight variations to these expressions, for example, applying the exponent outside of the parentheses ($(a_0 + \eta)^4$) may be useful to explore closer to the stability threshold.

The best algebraic expression for the particle flux identified in this work (un-normalized):
\begin{equation}
\Gamma_1 =   \left [ \sqrt{\frac{m_e}{m_i}} n_{0e} c_s \frac{ \rho_s ^2}{ a^2} \right ] \eta^2 \left (-0.18 \omega_{Te} +0.56 \omega_{ne} \right )
\end{equation}

Further refinements may be expected as additional scenarios are explored.  For example, additional simulation data at the pedestal top would be informative. 
Several applications of these models are envisioned, including: (1) rapid analysis of experimental discharges, (2) a component of more comprehensive models for pedestal structure, and (3) a complement to edge GK codes.

{\em Acknowledgements.--} We would like to thank Anthony Field for useful discussions.  This research used resources of the National Energy Research Scientific Computing Center, a DOE Office of Science User Facility, and the Texas Advanced Computing Center (TACC) at The University of Texas at Austin.  This work was supported by U.S. DOE Contract No. DE-FG02-04ER54742 and U.S. DOE Office of Fusion Energy Sciences Scientific Discovery through Advanced Computing (SciDAC) program under Award Number DE-SC0018429.  

\section{References}

\bibliography{my_refs}{}
\bibliographystyle{unsrt}

\appendix

\section{Data set}
\label{appendixA}

The parameters in the table are $\omega_{Te}=\frac{a}{L_{Te}} = 1/T_e d\rho_{tor} / dT_e$, $\omega_{ne}=\frac{a}{L_{ne}} = 1/n_e d\rho_{tor} / dn_e$, $\eta=\omega_{Te}/\omega_{ne}$, $\tau = \frac{T_{e0}}{T_{i0}}Z_{eff}$, the ratio of thermal to magnetic energy $\beta = 8 \pi n_{e0} T_{e0} /B_0^2$ (cgs), magnetic shear $\hat{s} =  \frac{\rho_{tor}}{q} \frac{d q}{d \rho_{tor}}$, the Debye length normalized to the sound gyroradius $\lambda_D/\rho_s$, and the normalized electron collision frequency $\nu^*_e = \frac{16}{3 \sqrt{\pi}} \frac{q Z^2}{\epsilon^{3/2}} \frac{R}{a}\frac{n_i}{n_e} \frac{\pi ln(\Lambda) e^4 n_{e0} a}{2^{3/2} T_{e0}^2}$.  These are standard GENE definitions, which can be found in the GENE documentation [genecode.org]. 

\begin{table}[htb!]
\label{tab:data}
\tiny
    \begin{tabular}{llllllllllll}
    Case & $\omega_{Te}$ & $\omega_{ne}$ & $\eta$ & $\tau$ & $\hat{s}$ & $\beta$  & $\nu_{*e}$ & $\lambda_D^2$ & $\rho_{tor}$ & Spec. & $Q_e/Q_{GB}$ \\
    1    & 26.7          & 5.97          & 4.46   & 0.975  & 4.06      & 0.00172  & 3.84       & 0.000208      & 0.97         & 1     & 107.0        \\
    2    & 31.2          & 8.34          & 3.74   & 1.72   & 2.24      & 0.00197  & 0.375      & 0.000409      & 0.965        & 1     & 61.9         \\
    3    & 49.4          & 13.7          & 3.6    & 1.58   & 3.04      & 0.00118  & 0.743      & 0.000455      & 0.975        & 1     & 116.0        \\
    4    & 19.4          & 2.91          & 6.67   & 1.17   & 3.68      & 0.00179  & 2.39       & 0.000238      & 0.97         & 1     & 244.0        \\
    5    & 47.6          & 11.0          & 4.33   & 0.87   & 5.92      & 0.001    & 5.85       & 0.000261      & 0.985        & 1     & 286.0        \\
    6    & 26.9          & 3.98          & 6.74   & 0.923  & 4.05      & 0.00139  & 3.15       & 0.000256      & 0.97         & 1     & 412.0        \\
    7   & 41.7          & 12.2          & 3.42   & 0.7    & 6.52      & 0.00074  & 8.11       & 0.000286      & 0.985        & 1     & 114.0        \\
    8   & 94.8          & 53.7          & 1.77   & 2.35   & 2.69      & 0.000895 & 0.27       & 0.000674      & 0.985        & 1     & 14.0         \\
    9   & 21.0          & 4.35          & 4.83   & 1      & 1.29      & 0.00208  & 0.812      & 0.000311      & 0.9675       & 1     & 33.1         \\
    10   & 52.9          & 18.1          & 2.93   & 1      & 1.12      & 0.00118  & 1.68       & 0.000351      & 0.98         & 1     & 49.0         \\
    11   & 74.2          & 12.9          & 5.75   & 2.8    & 4.99      & 0.000319 & 2.31       & 0.00112       & 0.5          & 1     & 938.0        \\
    12   & 19.1          & 8.83          & 2.16   & 2.35   & 0.909     & 0.00417  & 0.0661     & 0.000393      & 0.965        & 1     & 2.58         \\
    13  & 48.7          & 17.2          & 2.83   & 1.8    & 3.28      & 0.000888 & 0.816      & 0.000544      & 0.975        & 1     & 34.9         \\
    14  & 8.55          & 5.45          & 1.57   & 1      & 4.07      & 0.0029   & 0.28       & 0.000217      & 0.965        & 1     & 0.38         \\
    15  & 32.2          & 13.1          & 2.46   & 3.7    & -2.19     & 0.00299  & 0.117      & 0.000595      & 0.9675       & 1     & 4.28         \\
    16  & 30.1          & 20.6          & 1.46   & 0.664  & 6.72      & 0.000785 & 0.942      & 0.000408      & 0.972        & 1     & 0.807        \\
    17   & 41.5          & 25.5          & 1.63   & 0.556  & 13.4      & 0.000448 & 1.55       & 0.000506      & 0.982        & 1     & 2.1          \\
    18   & 22.6          & 10.9          & 2.08   & 0.666  & 2.17      & 0.000817 & 1.07       & 0.000873      & 0.962        & 1     & 7.92         \\
    19   & 29.8          & 13.1          & 2.29   & 0.623  & 3.35      & 0.000551 & 1.61       & 0.000989      & 0.972        & 1     & 14.0         \\
    20   & 32.3          & 13.5          & 2.4    & 0.581  & 6.32      & 0.000353 & 2.68       & 0.00113       & 0.982        & 1     & 19.8         \\
    21   & 25.5          & 17.4          & 1.47   & 1.7    & 3.37      & 0.00224  & 0.36       & 0.000239      & 0.975        & 1     & 1.72         \\
    22   & 28.3          & 15.3          & 1.85   & 1.7    & 3.37      & 0.00224  & 0.36       & 0.000239      & 0.975        & 1     & 6.13         \\
    23   & 31.1          & 13.2          & 2.36   & 1.7    & 3.37      & 0.00224  & 0.36       & 0.000239      & 0.975        & 1     & 15.9         \\
    24   & 33.9          & 11.1          & 3.07   & 1.7    & 3.37      & 0.00224  & 0.36       & 0.000239      & 0.975        & 1     & 35.2         \\
    25   & 25.8          & 5.82          & 4.43   & 1      & 1.12      & 0.00202  & 0.897      & 0.000301      & 0.97         & 1     & 39.8         \\
    26   & 52.9          & 18.1          & 2.92   & 1      & 1.12      & 0.00118  & 1.69       & 0.000351      & 0.98         & 1     & 50.5         \\
    27   & 42.5          & 14.9          & 2.85   & 2.0    & -0.789    & 0.00307  & 0.151      & 0.000808      & 0.975        & 1     & 22.5         \\
    28   & 48.3          & 20.2          & 2.39   & 2.0    & 1.33      & 0.0015   & 0.415      & 0.00093       & 0.978        & 1     & 17.1         \\
    29   & 42.5          & 15.0          & 2.84   & 2.0    & -0.715    & 0.00301  & 0.15       & 0.00082       & 0.975        & 1     & 12.5         \\
    30   & 42.5          & 18.4          & 2.31   & 2.0    & -0.895    & 0.00293  & 0.143      & 0.000848      & 0.975        & 1     & 6.91         \\
    31   & 48.0          & 16.6          & 2.9    & 2.0    & -0.477    & 0.0018   & 0.369      & 0.000447      & 0.978        & 1     & 20.3         \\
    32   & 37.5          & 24.7          & 1.52   & 2.35   & 1.04      & 0.00266  & 0.0991     & 0.00046       & 0.975        & 1     & 2.45         \\
    33   & 42.5          & 14.9          & 2.85   & 2.0    & -0.794    & 0.00307  & 0.15       & 0.000309      & 0.975        & 1     & 16.9         \\
    34   & 32.4          & 13.3          & 2.44   & 1      & -2.05     & 0.00296  & 0.12       & 0.000593      & 0.9675       & 1     & 10.2         \\
    35   & 32.4          & 13.3          & 2.44   & 2.5    & -2.05     & 0.00296  & 0.12       & 0.000593      & 0.9675       & 1     & 5.48         \\
    36   & 40.6          & 18.1          & 2.24   & 1.8    & 3.32      & 0.00121  & 1.58       & 0.000348      & 0.975        & 1     & 15.4         \\
    37   & 30.6          & 9.98          & 3.07   & 1.8    & 2.75      & 0.0019   & 1.04       & 0.000297      & 0.97         & 1     & 22.0         \\
    38   & 41.2          & 11.2          & 3.69   & 1.8    & 2.86      & 0.00152  & 0.605      & 0.000412      & 0.97         & 1     & 60.7         \\
    39   & 49.4          & 11.2          & 4.43   & 1.8    & 2.86      & 0.00152  & 0.605      & 0.000412      & 0.97         & 1     & 159.0        \\
    40   & 32.3          & 13.2          & 2.44   & 2.8    & -2.14     & 0.00408  & 0.08       & 0.000549      & 0.9675       & 1     & 4.15         \\
    41   & 32.2          & 13.1          & 2.46   & 3.7    & -2.25     & 0.0041   & 0.0784     & 0.000551      & 0.9675       & 1     & 3.39         \\
    42   & 42.5          & 14.9          & 2.86   & 2.0    & -0.811    & 0.00307  & 0.15       & 0.000402      & 0.975        & 1     & 16.1         \\
    43   & 48.4          & 17.2          & 2.81   & 1      & -0.0149   & 0.00161  & 0.355      & 0.000488      & 0.978        & 1     & 40.3         \\
    44   & 14.2          & 2.66          & 5.36   & n/a      & 2.51      & 0.00174  & 1.57       & 0.000214      & 0.96         & 3     & 36.0         \\
    45   & 6.84          & 1.37          & 5.0    & n/a      & 2.05      & 0.00415  & 0.249      & 0.000256      & 0.934        & 3     & 13.5         \\
    46   & 15.2          & 1.99          & 7.65   & n/a      & 3.02      & 0.0014   & 2.86       & 0.0002        & 0.973        & 3     & 60.2         \\
    47   & 42.7          & 15.9          & 2.69   & n/a      & 3.86      & 0.0015   & 0.993      & 0.000317      & 0.973        & 3     & 27.7         \\
    48   & 25.3          & 6.29          & 4.02   & n/a      & 2.87      & 0.00146  & 1.39       & 0.000263      & 0.964        & 3     & 45.7         \\
    49   & 33.9          & 10.1          & 3.35   & n/a      & 3.69      & 0.000966 & 4.34       & 0.000214      & 0.986        & 2     & 54.5         \\
    50   & 11.6          & 2.34          & 4.98   & n/a      & 1.91      & 0.00368  & 0.298      & 0.000261      & 0.945        & 3     & 15.0         \\
    51   & 29.9          & 9.48          & 3.16   & 1.72   & 2.24      & 0.00197  & 0.383      & 0.000406      & 0.965        & 1     & 32.9         \\
    52   & 47.5          & 15.5          & 3.06   & 1.58   & 3.04      & 0.00119  & 0.725      & 0.000459      & 0.975        & 1     & 58.7         \\
    53   & 62.2          & 23.1          & 2.69   & 1.41   & 4.45      & 0.000563 & 1.79       & 0.000556      & 0.985        & 1     & 71.6         \\
    54   & 15.9          & 8.83          & 1.8    & 2.35   & 0.909     & 0.00417  & 0.0661     & 0.000393      & 0.965        & 1     & 1.09         \\
    55   & 46.8          & 24.7          & 1.9    & 2.35   & 1.04      & 0.00266  & 0.0991     & 0.00046       & 0.975        & 1     & 8.24         \\
    56   & 23.1          & 10.3          & 2.24   & n/a      & 3.69      & 0.00268  & 3.36       & 0.000206      & 0.5          & 3     & 8.31         \\
    57   & 25.5          & 17.4          & 1.47   & n/a    & 3.37      & 0.00224  & 0.36       & 0.000239      & 0.975        & 2     & 2.4          \\
    58   & 28.3          & 15.3          & 1.85   & n/a    & 3.37      & 0.00224  & 0.36       & 0.000239      & 0.975        & 2     & 7.84         \\
    59   & 33.9          & 11.1          & 3.07   & n/a    & 3.37      & 0.00224  & 0.36       & 0.000239      & 0.975        & 2     & 40.7         \\
    60   & 26.9          & 3.98          & 6.74   & n/a      & 4.05      & 0.00139  & 3.15       & 0.000256      & 0.97         & 2     & 322.0        \\
    61   & 49.4          & 13.7          & 3.6    & n/a   & 3.04      & 0.00118  & 0.743      & 0.000455      & 0.975        & 2     & 139.0       
    \end{tabular}
    \end{table}

\cleardoublepage

\end{document}